\documentclass[10pt]{article}
\usepackage{graphicx}
\usepackage{latexsym}
\usepackage{amssymb}
\usepackage{amsfonts}
\usepackage{amsmath}
\usepackage{theorem}
\newtheorem{theorem}{Theorem}

\newtheorem{definition}{Definition}

\newtheorem{proposition}{Proposition}
\makeatletter
\newcommand{\contraction}[5][1ex]{%
  \mathchoice
    {\contraction@\displaystyle{#2}{#3}{#4}{#5}{#1}}%
    {\contraction@\textstyle{#2}{#3}{#4}{#5}{#1}}%
    {\contraction@\scriptstyle{#2}{#3}{#4}{#5}{#1}}%
    {\contraction@\scriptscriptstyle{#2}{#3}{#4}{#5}{#1}}}%
\newcommand{\contraction@}[6]{%
  \setbox0=\hbox{$#1#2$}%
  \setbox2=\hbox{$#1#3$}%
  \setbox4=\hbox{$#1#4$}%
  \setbox6=\hbox{$#1#5$}%
  \dimen0=\wd2%
  \advance\dimen0 by \wd6%
  \divide\dimen0 by 2%
  \advance\dimen0 by \wd4%
  \vbox{%
    \hbox to 0pt{%
      \kern \wd0%
      \kern 0.5\wd2%
      \contraction@@{\dimen0}{#6}%
      \hss}%
    \vskip 0.5ex
    \vskip\ht2}}

\newcommand{\contraction@@}[3][0.05em]{%
  \hbox{%
    \vrule width #1 height 0pt depth #3%
    \vrule width #2 height 0pt depth #1%
    \vrule width #1 height 0pt depth #3%
    \relax}}
\makeatother

\def\be{\begin{equation}}
\def\ee{\end{equation}}
\def\bc{\begin{center}}
\def\ec{\end{center}}

\usepackage{theorem}
\theorembodyfont{\upshape}

\begin{document}

\title{{Irreducible free energy expansion and overlaps locking in mean field spin glasses}}
\author{Adriano Barra\footnote{King's College London,
Department of Mathematics, Strand, London WC2R 2LS, United Kingdom,
and Dipartimento di Fisica, Universit\`a di Roma ``La Sapienza''
Piazzale Aldo Moro 2, 00185 Roma, Italy,
{\tt<Adriano.Barra@roma1.infn.it>}}}

\maketitle

\begin{abstract}
  Following the works of \cite{9},\cite{11}, we
  introduce a diagrammatic formulation for a
  cavity field expansion around the critical
  temperature.
  This approach allows us to obtain a theory for the overlap's fluctuations and,
  in particular, the linear part of the Ghirlanda-Guerra
  relationships (GG) (often called Aizenman-Contucci polynomials (AC)) in a very simple way.
  We show moreover how these constraints
  are $"$superimposed$"$ by the symmetry of the
  model with respect to the restriction required by thermodynamic
  stability.
  Within this framework it is possible to expand the free energy in
  terms of these irreducible overlaps fluctuations and in a form that simply put in evidence
  how the complexity of the solution is related to the complexity of the entropy.
\end{abstract}

\bigskip

\bc\textbf{key words:} Cavity field, Ghirlanda-Guerra, stochastic stability, Aizenman-Contucci polynomials.\ec

\bigskip

\section{Introduction}

The configuration space of the Sherrington-Kirkpatrick model (SK)
is built by N Ising variables $\left\{ \sigma _{i}\right\} $
 which interact
respecting the following Hamiltonian $H(\{\sigma\})$ \cite{1}\cite{2}\cite{3}:

\be\label{proviamoci}
H(\{\sigma\})=-\frac{1}{\sqrt{N}}%
\sum^{N,N}_{i<j}J_{ij}\sigma _{i}\sigma
_{j}-h\sum\limits_{i}^{N}\sigma _{i} \ee
where h is an external scalar field, which, for the rest of the paper, we will set to zero, and $J_{ij}$ are i.i.d.
Gaussian variables such that $P\left[
J_{ij}\right] $:

\be\label{P}
P[J_{ij}]= \prod_{i<j}[{\frac{1}{\sqrt{2\pi}}}\exp \left( -\frac{1}{2}%
J^2_{ij}\right) ]. \ee

Once defined the partition function $Z(\beta, J_{ij})$ as

\be\label{Z}
Z_{N}(\beta,J_{ij})=\sum_{\{\sigma\}}e^{\frac{\beta}{\sqrt N}
\sum_{ij}J_{ij}\sigma_i\sigma_j}\ee
with the inverse temperature $\beta =T^{-1}$ in proper units, the
Gibbs measure $G_N(\sigma; \beta, J_{ij})$
$$
G_N(\sigma)= \frac{1}{Z_N} \exp^{-\beta H_N (\sigma)}
$$
and all the standard statistical mechanics package, let
us introduce also the average
over the $\left\{ J_{ij}\right\} $ denoting by \textbf{E}(.):
\begin{center}
\be\label{E}
\textbf{E}\left[ f\left( J_{ij}\right) %
\right] =\prod_{i<j}{\frac{1}{\sqrt{2\pi}}}\int dJ_{ij}f\left(
J_{ij}\right) \exp (-J_{ij}^{2}/2). \ee
\end{center}
We are interested in the explicit
expression for the quenched average of the free energy density (we will use the symbol $F$ for the free energy and $f$ for its density):
\be \alpha(\beta)= lim_{N \rightarrow \infty} \alpha_N(\beta)=-\beta
\mathbf{f}\left( \beta\right) =lim_{N \rightarrow \infty}
N^{-1}\mathbf{E}\ln Z_{N}\left( \beta ,J_{ij}\right). \ee
Let us introduce some basic notation:
For a  function $f(\{\sigma_i\}_{(i=1,..,N)},J)$ of the degrees of freedom $\{\sigma\}$ and the noise $J$ we define:
\bigskip

$\omega[f(\{\sigma\})] =
\frac{\sum_{\{\sigma\}}f(\{\sigma\})e^{-\beta
H(\{\sigma\})}}{\sum_{\{\sigma\}}e^{-\beta H(\{\sigma\})} } $
\
as the thermal average,

\bigskip

$\Omega[f(\{\sigma^{\alpha})\}] = \omega_1[f(\sigma^1)] \cdot
\omega_2[f(\sigma^2)] \cdot ... \cdot \omega_n[f(\sigma^n)] $ \ as
the generalized thermal average over the replicas of the system
\cite{5},

\bigskip

$\langle f(\{\sigma\},J)\rangle = E\Omega[f\{\sigma\}]$ as the
average first on the thermal weight and than over the coupling
\cite{5},

\bigskip

$q_{\alpha\beta}=N^{-1}\sum_{i}^{N}\sigma_i^{\alpha}\sigma_i^{\beta}$
as the overlap of two replicas,

\bigskip

$q_{\alpha\beta\gamma}=N^{-1}\sum_{i}^{N}\sigma_i^{\alpha}\sigma_i^{\beta}\sigma_i^{\gamma}$
\
as the overlap over three replicas and so on.

\section{Cavity field: toward a diagrammatic formulation}

The main idea of the {\em cavity field} method is to look for an
explicit expression of  $\alpha(\beta)=-\beta f(\beta)$ upon
increasing the size of the system from $N$ particles (the cavity) to
$N+1$ (see \cite{3}\cite{9}\cite{10}\cite{13}) so that, in the limit
of N that goes to infinity

\bc $[-\beta F_{N+1}(\beta)]-[-\beta F_N(\beta)]=-\beta f(\beta) +
O(\frac{1}{N})$ \ec

because the existence of the thermodynamic limit \cite{14} implies only  vanishing correction
of the free energy density.

\bigskip

So, following \cite{11} in a variant nowadays called stochastic stability \cite{12},
let us introduce an extended partition function \cite{10}
able to manage the interaction with the added spin through a
control parameter $t \in [0,\beta^2]$ such that for t=0 we have
the classical partition function of $N$ spins while for
$t=\beta^2$ we get the expression of the partition function for the
larger system with a little temperature shift which vanishes in
the thermodynamic limit.

\bigskip
\be\label{cf10} Z_{N,t}= \sum_{\sigma_1,..,\sigma_N}e^{-\beta
H_N(\sigma, J)+\sqrt{\frac{t}{N}}\sum^N_{i}J_i\sigma_i} \ee
when $t=\beta^2$, redefining $J_i \rightarrow J_{i,N+1}$
and making the transformation $\sigma_i \rightarrow
\sigma_i\sigma_{N+1} \ \forall i$ we obtain the partition function
for a system of $N+1$ spin at a scaled temperature $\beta^*$ such that

\be\label{cfx} \beta^*=\beta \sqrt{(N+1)/N} \rightarrow \beta
\mbox{ for } N \rightarrow \infty. \ee
Now we will try to formalize some concepts which will be useful developing  this  version of cavity method:

\begin{proposition}{\itshape
The averages $ \langle.\rangle$ are invariant under replica
symmetry.} In fact if we consider a generic element g of the group
\textbf{G} of the permutation of  s replicas, defining a generic
replica (a) and its transformed (a'), the action of g on (a) is such
that:
$$
\textbf{G}\ni g:g*(a)=(a')\Rightarrow \langle F(q_{ab})\rangle=
\langle F(q_{AB'})\rangle
$$
\end{proposition}
\begin{proposition}{\itshape
The averages $ \langle .\rangle$  are invariant under gauge
symmetry:} \be \sigma_i^a \rightarrow \epsilon_a \sigma_i^a \ee \be
q_{ab}\rightarrow  \epsilon_a \epsilon_b q_{ab} \ee being
$\epsilon_a=\pm1$
\end{proposition}
This symmetry is a consequence of the parity of the state $\Omega$
and of dichotomy of Ising variables.
\begin{definition}{\itshape
we define as filled a polynomial of the overlaps in which
every replica appears an even number of times.}
\end{definition}
\begin{definition}{\itshape
we define as fillable a polynomial in which the above property is
obtainable by multiplying the polynomial itself for just one overlap of two replicas
that is exactly the  needed term to fill
the expression.}
\end{definition}
\begin{definition}{\itshape
we define as un-fillable a polynomial which is neither filled nor
fillable.}
\end{definition}

Example: the following expressions are filled: $q_{12}^2$, \
$q_{12}q_{23}q_{31}$.

Example: the following expression are fillable: $q_{12}$, \
$q_{12}q_{23}$.

Example: the following expressions are un-fillable: $q_{1234}$, \
$q_{12}q_{23}q_{45}$.
\begin{definition}{\itshape
We define the {\em cavity function}  $\Psi(t)$ as:}

\be \label{cf12}
\Psi(t)=\textbf{E}[\ln\omega(e^{\frac{\sqrt{t}}{{N}}\sum_iJ_i\sigma_i})]
=\textbf{E}[\ln\frac{Z_{N,t}}{Z_N}] \ee
\end{definition}
and let F be a generic function of the $\{\sigma\}$ such that,
\begin{definition}{\itshape
we define the {\em generalized Boltzmann state} obtained using the
partition function \cite{10} as:}
\be \label{cf0}
\omega_t(F)=\frac{\omega(Fe^{\sqrt{\frac{t}{N}}\sum_i^NJ_i\sigma_i})}
{\omega(e^{\sqrt{\frac{t}{N}}\sum_i^NJ_i\sigma_i})}. \ee
\end{definition}
\begin{theorem}
{\itshape In the $N \rightarrow \infty$ limit  the averages $\langle
.\rangle$ of the filled polynomials are t-independent in $\beta$
average.}
\end{theorem}

\textbf{Proof}

Without loss of generality we will prove the theorem in the simplest
case (for $q^2_{12}$). Let us write the cavity function as
\be\label{cf13} \Psi(t)=\textbf{E}[lnZ_{Nt}]-\textbf{E}[lnZ_{N}] \ee
and derive it respect to $\beta$:
\be \label{cf16} \frac{d\Psi(t)}{d\beta}=\frac {\beta N}{2}(\langle
q_{12}^2\rangle - \langle q_{12}^2\rangle_t). \ee We can introduce
an auxiliary function $\Upsilon_N(t,\beta)$:
\be
\Upsilon_N(t,\beta)=\frac{4}{N}d_{\beta^2}[\Psi_N(t,\beta)]
\ee
and integrate it in the interval $[\beta_1,\beta_2]$:
\be \int_{\beta_1^2}^{\beta_2^2}\Upsilon_N(t,\beta)d\beta^2=
\frac{4}{N}[\Psi_N(t,\omega(\beta_2))-\Psi_N(t,\omega(\beta_1))].
\ee
In the thermodynamic limit $\Psi(t)$ remains limited and the
second member goes to zero; so,  $\forall$ [$\beta_1^2$,$\beta_2^2$]
we can always extract a subsequence such that the $\Upsilon_N(t)$
converge to zero in  Lebesgue measure.

\bigskip

The next theorem is crucial for this paper, so, for the sake of simplicity we divided it in two part: the first one will be the following lemma and it will make us able to
proof the theorem itself which will be showed immediately after.

\bigskip

\textbf{Lemma 1}
{\itshape Let  $\omega(.)$ and $\omega_t(.)$ be the states defined
respectively by the canonical partition function and by the
extended one; if we consider the ensemble of index  $\{i_1,..,i_r\}$
with $r\in [1,N]$, then for $t=\beta^2$ the following relation holds:}
\be
\omega_{N,t=\beta^2}(\sigma_{i_1},...,\sigma_{i_r})=
\omega_{N+1}(\sigma_{i_1},...,\sigma_{i_r},\sigma_{N+1}^r)
+O(\frac{1}{N} )
\ee
{\itshape where r is an exponent, not a replica index, so if r is
even $\sigma_{N+1}^r=1$, while if is odd
$\sigma_{N+1}^r=\sigma_{N+1}$.}

\bigskip

\textbf{Proof}

Let us write the $\omega_t$ for $t=\beta^2$ defining for the sake
of simplicity $\sigma=\sigma_{i_1}...\sigma_{i_r}$:

\be
\omega_{N,t=\beta^2}(\sigma)=\textbf{E}[\sum_{\{{\sigma}\}}\frac{1}{Z_{N,
\beta^2}}e^{\frac{\beta}{\sqrt{N}}\sum_{i<j}J_{ij}\sigma_i\sigma_j +
\frac{\beta}{\sqrt{N}}\sum_i J_i\sigma_i}\sigma]. \ee
Introducing a sum over $\sigma_{N+1}$ at the numerator and
at the denominator, (which is the same as multiply and divide for $2^N$
because there is no dependence to $\sigma_{N+1}$) and
making the transformation  $\sigma_i\rightarrow
\sigma_i\sigma_{N+1}$, the variable $\sigma_{N+1}$ appears at the
numerator and it is possible to build the status at $N+1$
particles.
Expanding $\beta^*$ for large N we have that:

\bc \be \omega_{N,t= \beta^2}(\sigma)=
\omega_{N+1}(\sigma\sigma_{N+1}^r) +O(\frac{1}{N}). \ee \ec
Using this lemma we are able to proof the following:

\bigskip

\begin{theorem}
{\itshape Let $Q_{ab}$ be a fillable polynomial of the overlaps,
(this means that  \ $\langle q_{ab}Q_{ab}\rangle$ is filled). We
have:}
\be
\lim_{N \rightarrow
\infty}\lim_{t\rightarrow\beta^2}\langle Q_{ab}\rangle_t= \langle
q_{ab}Q_{ab}\rangle
\ee
\end{theorem}

\textbf{Proof}

We write the average splitting the dependence from the non filled
replicas a,b to the others:
\be\label{101} Q_{ab}=\sum_{ij}\frac{\sigma_i^a\sigma_j^b}{N^2}
Q_{ij}(\sigma). \ee
We have indicated with $Q_{ij}(\sigma)$  the product of the
non filled replicas. Factorizing  the state $\Omega$ we obtain:

\be \langle Q_{ab}\rangle_t =
\frac{1}{N^2}\textbf{E}[\sum_{ij}\Omega_t(\sigma_i^a\sigma_j^b
Q_{ij}(\sigma))]= \ee

\be\label{102}
= \frac{1}{N^2}\textbf{E}[\sum_{ij}
\omega_t(\sigma_i^a)\omega_t(\sigma_j^b)\Omega_t(Q_{ij})].
\ee
Now we write the last expression for $t=\beta^2$; using the lemma,
the states acting on the replicas $a$ and $b$ are
\be\label{103} \omega_{t=\beta^2}(\sigma_i^a)=\omega
(\sigma_i^a\sigma_{N+1}^a) + O(\frac{1}{N}) \ee
while the remaining product state  $\Omega_t$ continue to work on
a even number of replicas and is not modified
\be\label{104} \Omega_{t=\beta^2}(Q_{ij})=\Omega(Q_{ij}). \ee
Putting all the replicas in  a unique product state we
have:
\bc \be\label{104} \omega(\sigma_i^a\sigma_{N+1}^a)
\omega(\sigma_i^b\sigma_{N+1}^b)\Omega(Q_{ij})=
\Omega(\sigma_i^a\sigma_j^b \sigma_{N+1}^a\sigma_{N+1}^b Q_{ij}).
\ee \ec
Using replica symmetry we can write the index $N+1$ as a $"k"$
dumb one; pass on, we can sum on all "k" from 1 to N and divide
for N because the terms with k equal to one of the index i,j are of
order $O(\frac{1}{N})$ and became irrelevant in the $N \rightarrow
\infty$ limit. So:
\be\label{106} \langle Q_{ab}\rangle_{\beta^2}= N^{-3}\textbf{E}[
\sum_{ijk}\Omega(\sigma_k^a \sigma_k^b \sigma_i^a
Q_{ij}\sigma_j^b)]+O(\frac{1}{N}) \ee and in the thermodynamic limit
we have the proof.

\bigskip

Let us remember (this will be useful soon) an important theorem, due
to Guerra, stating that the streaming equations for the cavity
fields are \cite{6}:
\begin{theorem}
{\itshape Let $F_s \in {\{A_s\}}$, where ${\{A_s\}}$ is the
algebra built by the overlaps of s replicas; the following
streaming equation for F holds:}
\be\label{108}
\partial_t \langle F_s\rangle_t =<F_s (\sum_{a,b}q_{a,b}
-s\sum_{a=1}^sq_{a,s+1}+ \frac{s(s+1)}{2}q_{s+1,s+2})\rangle_t. \ee
\end{theorem}
By now we define abstract graphs in this way: We introduce numbered
vertexes to label replicas of the system and lines between them to
label their overlaps; for example $\langle q_{12}\rangle_t$ will be
associated to the graph 1\includegraphics{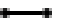}2, $\langle
q_{12}^2\rangle_t$ to 1\includegraphics{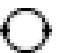}2 and so on. It is
obvious that there is a one to one connection between polynomials
and graphs.

\section{Free Energy and Cavity Function}

Now we want to show how it is possible to expand the SK free
energy via these graphs.
Look carefully at the energy of the system, using thermodynamic
relations we have:
\be
\textbf{E}[\omega(H_N)]=-N\partial_{\beta}\alpha_N(\beta)
=-\textbf{E}[\frac{\partial_{\beta}Z_N(\beta,J)}{Z_N(\beta,J)}]=
\frac{-N\beta}{2}(1- \langle q_{12}^2\rangle).
\ee
%
So in our vocabulary:
%
%
\begin{proposition}
{\itshape The internal energy of SK model is:}
$$
\textbf{E}[u_N(\beta)]=\frac{1}{N}\textbf{E}[\omega_J(H_N(\sigma,J))]=
-\frac{\beta}{2}(1-\langle \includegraphics{02a.eps}\rangle).
$$
\end{proposition}

\begin{theorem}
{\itshape Assuming that }
$lim_{N\rightarrow\infty}\Psi_N(\beta)=\Psi(\beta)$
{\itshape exists uniformly in every compact} $0\leq \beta \leq
\beta^*${\itshape than the following relation holds}:
\be\label{215} \alpha(\beta)+\frac {\beta}{2}\alpha'_{\beta}=
ln2+\Psi(\beta) \ee
\itshape{where $\Psi_N(\beta)$ is the cavity function previously
defined}.
\end{theorem}
This theorem give us a connection between the cavity function in
the limit $t\rightarrow \beta^2$, the free energy and the internal
energy of the system.

\bigskip

\textbf{Proof:}

Consider the partition function of a system of $(N+1)$ spins and
point out with $\beta$ the effective temperature and with $\beta^*$
the scaled one:

\be\label{217} Z_{N+1} =\sum_{\{{\sigma_1,..,\sigma_{N+1}}\}}
e^{\frac{\beta}{\sqrt{N+1}}\Sigma_{i<j}^{N+1}J_{ij}\sigma_i\sigma_j}=
2\sum_{\{{\sigma_{N}}\}}
e^{\frac{\beta^*}{\sqrt{N}}\Sigma_{i<j}^{N}J_{ij}\sigma_i\sigma_j}
e^{\frac{\beta}{\sqrt{N+1}}\sum_i^N J_i \sigma_i}.
\ee
Multiplying and dividing for $Z_N(\beta^*)$, taking the logarithm,
subtracting from every members the
quantity $lnZ_{N+1}(\beta^*)$ and expanding
$lnZ_{N+1}(\beta)$ around $\beta= \beta^*$ we have:
\be
lnZ_{N+1}(\beta)-lnZ_{N+1}(\beta^*) = (\beta- \beta^*)
\partial_{\beta^*}lnZ_{N+1}(\beta^*)+ O(d\beta^2)
\ee
being:
\be
(\beta-\beta^*)=\beta^*(\sqrt{\frac{N+1}{N}}-1)=\frac{\beta^*}{2N} +
O(N^{-1}).
\ee
Substituting $\beta$ with $\beta^*$ inside the state $\omega$
apart corrections $O(N^{-1})$ we have:
\be lnZ_{N+1}(\beta^*)
+(\beta-\beta^*)\partial_{\beta^*}lnZ_{N+1}(\beta^*)=
\ee
\bc
$$
ln2+lnZ_N(\beta^*)+ln\omega_N^{(\beta^*)}(e^{\frac{\beta}{\sqrt{N+1}}
\sum_iJ_i\sigma_i}) +O(N^{-1}).
$$
\ec
Taking the average \textbf{E}, using the variable $\alpha$ and
renaming $\beta^* \rightarrow \beta$ in the thermodynamic limit we get:

\be \alpha(\beta) + \frac{\beta}{2}\partial_{\beta}\alpha(\beta) =
ln2+ \Psi(\beta). \ee
and this is the thesis of the theorem (see also
\cite{guerriero,baffioni} for further details concerning the theorem
above).
So we can study cavity function to understand properties
 of the free energy. To do this, derive now the cavity function using the
standard integration over Gaussian noise:

\be
\partial_t\Psi(t)=
=\frac{1}{2}\textbf{E}[1-\frac{1}{N}\sum_i\omega_t^2(\sigma_i)]=
\frac{1}{2}(1- \langle q_{12}\rangle_t). \ee
Note, being $Z_{N,0}=Z_N$, that the cavity function
itself is obtainable as an integral over t of the simplest
two-replicas overlap weighted with the generalized Boltzmann state
$\omega_t$.

\be\label{666} \Psi_N(t,\omega) -\Psi_N(0,\omega)= \Psi_N(t,\omega)=
\frac{1}{2}\int_0^t dt'(1- \langle q_{12}\rangle_{t'}). \ee
\bigskip
Let us show how to expand  $q_{12}=
\includegraphics{01a.eps}$:
As the parameter of expansion we define the number of lines of the
filled graphs; surely for the non-filled ones their parameter will be
the corresponding one of the same order filled graph.
Before starting the expansion we present just some simple example:

es.1)$\langle \includegraphics{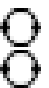}\rangle$ this graph is of the
fourth order, filled and means the quantity $\langle
q_{12}^2q_{34}^2\rangle$.

es.2)$ \langle \includegraphics{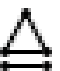}\rangle$ this graph is of
the fifth order, fillable and means the quantity $\langle
q_{12}q_{23}q_{13}q_{45}\rangle$ (moreover its filled corresponding
is $\langle \includegraphics{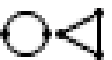}\rangle$).

\bigskip

Remembering  the streaming equation and the pull of rules
developed before we can write:
\be\label{g1}
\partial_t \langle \includegraphics{01a.eps}\rangle_t = \langle \includegraphics{02a.eps} -4
\includegraphics{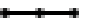}+3 \includegraphics{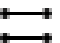}\rangle_t
\ee
At this first step we find a filled graph, $q_{12}^2$; For it the
perturbative expansion is ended here. Now we have to
express also the other graphs (the non filled ones) in terms of the filled:

\be\label{g2}
\partial_t \langle \includegraphics{03a.eps}\rangle_t=
\langle \includegraphics{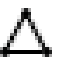} +2 \includegraphics{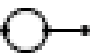}
-6\includegraphics{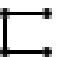}-3\includegraphics{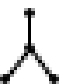} +6
\includegraphics{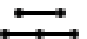}\rangle_t
\ee

\bigskip

\be\label{g3}
\partial_t \langle \includegraphics{04b.eps}\rangle_t= \langle 2\includegraphics{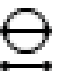}+
4 \includegraphics{04e.eps}-16\includegraphics{05d.eps}
+10\includegraphics{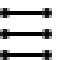}\rangle_t. \ee
\bigskip
Another filled one has been obtained  (the first of the r.h.s. of the eq.
(\ref{g2})). However we have to go on for the others.
$$
\partial_t \langle \includegraphics{04c.eps}\rangle_t=
$$
\be\label{g4} = \langle
\includegraphics{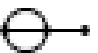}+\includegraphics{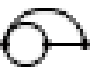}+
\includegraphics{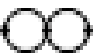}-3\includegraphics{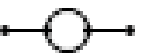} -3\includegraphics{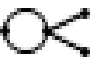}
-3\includegraphics{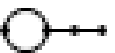}+6\includegraphics{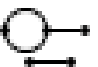}\rangle_t \ee
and so on.
We stop here our expansion for now, showing just the result for
higher orders.
Neglecting terms order $\beta^{10}$ we have:

\be\label{g12} \langle
\includegraphics{01a.eps}\rangle_t= \langle \includegraphics{02a.eps}t
-2\includegraphics{03b.eps}t^2
-\frac{4}{3}\includegraphics{04h.eps}t^3+
\includegraphics{04f.eps}t^3+6\includegraphics{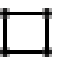}t^3-
\ee

$$
+10\includegraphics{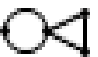}t^4
-\frac{20}{3}\includegraphics{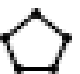}t^4
-8\includegraphics{nu01.eps}t^4\rangle.
$$
We can now write (formally at any level of knowledge) the
$\Psi(\beta)$ and, being known also the expression for the internal
energy, we are able to write the expression for the free energy.

\begin{proposition}
{\itshape the free energy expansion via irreducible overlaps
fluctuations is} \bc \be\label{g14} \alpha(\beta)= \langle ln2
+\frac{\beta^2}{4}[1+(1-\beta^2)\includegraphics{02a.eps}]
+\frac{\beta^6}{3}\includegraphics{03b.eps}+
\frac{\beta^8}{6}\includegraphics{04h.eps}-
 \ee
$$
\frac{\beta^8}{8}\includegraphics{04f.eps}
-\frac{3\beta^8}{4}\includegraphics{04d.eps}
-\beta^{10}\includegraphics{05b.eps}
+\frac{12\beta^{10}}{5}\includegraphics{05m.eps}
+\frac{2\beta^{10}}{3}\includegraphics{nu01.eps}\rangle+O(\beta^{12}).
$$
\ec
\end{proposition}

As we can immediately see this solution is built by more and more overlaps correlation functions (complete graphs) which a replica-symmetric theory cannot generate; moreover it is also easy to see that  in the high temperature  phase it reproduces the right expression and that the birth of all these correlations functions is just due to the entropy, being the energy density:

\be -\partial_{\beta}\alpha(\beta)=-\frac{\beta}{2}(1- \langle
\includegraphics{02a.eps}\rangle) \ee

while the entropy is:

\be S(\beta)= \langle \ln2
-\frac{\beta^2}{4}(1-3\includegraphics{02a.eps})-\frac{\beta^4}{4}\includegraphics{02a.eps}+\frac{\beta^6}{3}\includegraphics{03b.eps}+
\frac{\beta^8}{6}\includegraphics{04h.eps}-
\frac{\beta^8}{8}\includegraphics{04f.eps}\rangle+... \ee

\section{Overlap Constraint Generator}

Now we want to use a pure thermodynamical approach to obtain the
Aizenman-Contucci polynomials: we simply impose on the theory that
the total derivative of the free energy with  respect to $\beta^2$
has to be (in modulus) the internal energy of the system: (doing
this we use the redundant formalism of $\alpha(\beta,\langle
.\rangle_{\beta^2})$ instead of $\alpha(\beta)$ to put in evidence
how to perform the right derivative.)

\be \alpha(\beta,\langle .\rangle_{\beta^2})=[ln2+\Psi(t,\langle
.\rangle_t)-\frac{t}{4}(1- \langle q_{12}^2\rangle)]_{t=\beta^2}.
\ee
Its total derivative respect to $\beta^2$ is:
\be \frac{d}{d\beta^2}\alpha(\beta,\langle .\rangle_{\beta^2})
=\partial_{\beta^2}\alpha(\beta,\langle
.\rangle_{\beta^2})+\sum_{\langle .\rangle}
\frac{\partial\alpha(\beta,\langle
.\rangle_{\beta^2})}{\partial\langle .\rangle} \frac{\partial\langle
.\rangle}{\partial\beta^2}, \ee
where the expression $\sum_{\langle .\rangle}$ means {\em the sum
over all the filled graphs
 contained in the} $\alpha$ and this derivative is equal (in modulus) to the internal energy:
\be \frac{d}{d\beta^2}\alpha(\beta,\langle .\rangle_{\beta^2})=
\frac{1}{2\beta}\alpha'_{\beta}=\frac{1}{4}(1-\langle
q_{12}^2\rangle). \ee
Start to calculate the total derivative: the partial one is
\be
\partial_{\beta^2}\alpha(\beta,\langle .\rangle_{\beta^2})=[\partial_t\Psi(t,\langle .\rangle)-
\frac{1}{4}(1- \langle q_{12}^2\rangle)]_{t=\beta^2} \ee
where
\be
\partial_t\Psi(t,\langle .\rangle)_{t=\beta^2}=\lim_{t\rightarrow \beta^2}
\frac{1}{2}(1-\langle q_{12}\rangle_t)=\frac{1}{2}(1-\langle
q_{12}^2\rangle). \ee
So the partial and the total derivative are the same
\be \frac{d}{d\beta^2}\alpha(\beta,\langle .\rangle_{\beta^2})=
\partial_{\beta^2}\alpha(\beta,\langle .\rangle_{\beta^2})
\ee
therefore the following quantity has to be identically zero:
\be \sum_{\langle .\rangle} \frac{\partial\alpha(\beta,\langle
.\rangle_{\beta^2})}{\partial \langle .\rangle}
\frac{\partial\langle .\rangle}{\partial\beta^2}=0. \ee
Giving an explicit expression we have:

\begin{proposition}
{\itshape the following expansion is the $''$thermodynamic$''$ generator of the Aizenman-Contucci restrictions for the overlaps fluctuations:}
\end{proposition}
\be\label{salame}
\langle\frac{\partial\alpha}{\partial\includegraphics{02a.eps}}
\frac{\partial\includegraphics{02a.eps}}{\partial\beta^2}+
\frac{\partial\alpha}{\partial\includegraphics{03b.eps}}
\frac{\partial\includegraphics{03b.eps}}{\partial\beta^2}+
\frac{\partial\alpha}{\partial\includegraphics{04h.eps}}
\frac{\partial\includegraphics{04h.eps}}{\partial\beta^2}+
\frac{\partial\alpha}{\partial\includegraphics{04f.eps}}
\frac{\partial\includegraphics{04f.eps}}{\partial\beta^2}+
...\rangle=0. \ee

We try now to extrapolate some information from the above statement,
starting to note that the apparently redundant word {\itshape
thermodynamics} has been used to put in evidence the difference with
the next overlap constraint generator which will be called {\itshape
symmetric} because it will be derived just using the properties of
the symmetry of the model.
\newline
Starting to study every single monomial  we find a common
structure to all members of the polynomial.
For the first one:

\be \langle \frac{\partial\alpha}{\partial\includegraphics{02a.eps}}
\frac{\partial\includegraphics{02a.eps}}{\partial\beta^2} \rangle=
\frac{\beta^2}{4}(1-\beta^2)\frac{1}{2\beta}\partial_{\beta}
\textbf{E}\Omega_s(q_{12}^2)= \ee

\be =Nf(\beta)[\langle (q_{12}^2)
(\sum_{\alpha,\beta}q_{\alpha,\beta}^2-s\sum_{\alpha}q_{\alpha,s+1}^2
+\frac{s(s+1)}{2}q_{s+1,s+2}^2)\rangle]. \ee
For the second one:

\be \langle \frac{\partial\alpha}{\partial\includegraphics{03b.eps}}
\frac{\partial\includegraphics{03b.eps}}{\partial\beta^2} \rangle=
\frac{\beta^6}{3}\frac{1}{2\beta}\partial_{\beta}\textbf{E}\Omega_s
(q_{12}q_{13}q_{23})= \ee

\be =Ng(\beta)[\langle (q_{12}q_{13}q_{23})
(\sum_{\alpha,\beta}q_{\alpha,\beta}^2-s\sum_{\alpha}q_{\alpha,s+1}^2
+\frac{s(s+1)}{2}q_{s+1,s+2}^2)\rangle]. \ee
We are now able to write the (\ref{salame}) in this other way:

\be\label{pv1} \langle
[\frac{\beta^2}{4}(1-\beta^2)\includegraphics{02a.eps}+
\frac{\beta^6}{3}\includegraphics{03b.eps}+
\frac{\beta^8}{6}\includegraphics{04h.eps}+
\frac{\beta^8}{8}\includegraphics{04f.eps}-
\frac{3\beta^8}{4}\includegraphics{04d.eps}+...]\times \ee

$$
\times[\sum_{\alpha,\beta}q_{\alpha,\beta}^2-s\sum_{\alpha}q_{\alpha,s+1}^2
+\frac{s(s+1)}{2}q_{s+1,s+2}^2]\rangle=0.
$$
We started taking as the parameter of the expansion the number
of links inside a graph but, because there is  a complete
equivalence with the order of $\beta^2$, now we do the opposite: we
think as the parameter $\beta^2$ and being the equation
(\ref{salame}) a polynomial in $\beta^2$ equal to zero we have the
right  to put to zero every monomial inside.
The first order give us:

\be \langle \frac{\beta^2}{4}(1-\beta^2)\includegraphics{02a.eps}
[\sum_{\alpha,\beta}q_{\alpha,\beta}^2-s\sum_{\alpha}q_{\alpha,s+1}^2
+\frac{s(s+1)}{2}q_{s+1,s+2}^2]\rangle=0 \ee
If we put out from the averages the temperature function in order to
neglect it when it's a well defined quantity we obtain:

\be\label{df} \langle
\includegraphics{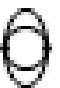}-4\includegraphics{04h.eps}+3
\includegraphics{04f.eps}\rangle=0
\ee
This is the first Aizenman-Contucci relation or the first linear part of the Ghirlanda-Guerra identities.
\
For the second order it's the same:
\be\label{GG2}
 \langle \includegraphics{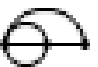}-3\includegraphics{05b.eps}+2
\includegraphics{nu01.eps}\rangle=0
\ee
Now we can start to look at the third order: we want to prove that
the restrictions imposed by thermodynamic arguments are $''$sweet$'$ with
respect to the ones showed using symmetry properties of SK model
Look at the expansion in $\beta^2$ of (\ref{pv1}), thermodynamic
requires that:

\be\label{succo} \langle [\frac{1}{6}\includegraphics{04h.eps}-
\frac{1}{8}\includegraphics{04f.eps}-
\frac{3}{4}\includegraphics{04d.eps}]
[\sum_{\alpha,\beta}q^2_{\alpha,\beta} -s\sum_{\alpha}q^2_{\alpha,
s+1}+\frac{s(s+1)}{2}q^2_{s+1,s+2}]\rangle=0. \ee
Anyway, reading again the pull of rules developed before, we could
state that:

\begin{proposition}
{\itshape the following expansion is the $''$symmetric$''$ generator of the Aizenman-Contucci restrictions for the overlaps fluctuations:}
\end{proposition}

\be \lim_{N\rightarrow\infty}\lim_{t \rightarrow
\beta^2}\partial_t[filled \ graph]=0 \ee
This property has been demonstrated implicitly because in the
expansion previously showed we did not develop the graphs $q^2_{12}$
,$q_{12}q_{23}q_{31}$ being them independent from the t parameter.
Now we can look at this property also like a generator of
constraints for the overlaps. Here we have some examples:
Start with the first filled graphs and find again the
linear Ghirlanda-Guerra relations:

\bc\be \lim_{N\rightarrow\infty}\lim_{t \rightarrow
\beta^2}\partial_t[q^2_{12}] = \langle
\includegraphics{04g.eps}-4\includegraphics{04h.eps}+
3\includegraphics{04f.eps}\rangle=0 \ee

\be \lim_{N\rightarrow\infty}\lim_{t \rightarrow \beta^2}\partial_t[q_{12}q_{23}q_{31}]=
 \langle \includegraphics{05i.eps}-3\includegraphics{05b.eps}+2
\includegraphics{nu01.eps}\rangle=0
\ee\ec
We can show how  the derivative and the t-limit act together (in
the thermodynamic limit) on a filled graph:

\bc\be \lim_{N\rightarrow\infty}\lim_{t \rightarrow
\beta^2}\partial_t \langle [filled \ graph]\rangle= \ee

$$
=\langle [filled \ graph][\sum_{\alpha,\beta}q^2_{\alpha,\beta}
-s\sum_{\alpha}q^2_{\alpha,
s+1}+\frac{s(s+1)}{2}q^2_{s+1,s+2}]\rangle.
$$\ec

It's interesting that such constrictions, that are
properties of SK model, are not strictly imposed by
thermodynamic and are stronger that that ones:
If we came back to look again at the fourth  order ($\beta^8$) in our expansion we have that
thermodynamic requires for the sum of the three filled graphs in
the expression (\ref{succo}) (multiplied by the second parenthesis
of the expression) to be zero but we have already show that
also every single term in the above expression have to be zero.
So the thermodynamic conditions are (respected and) weaker than that required by the symmetry of the model
and this has the consequence that the system is not allowed to exploit the whole space but
only the part of it where these restrictions hold.
Moreover we found in this way a manner to build explicitly a  kind of {\em constriction tree} for the overlap fluctuations because to obtain the restrictions at the fourth order we had to derive the unique graph of the second order and, after the limit of t that goes to $\beta^2$, we had to put that expression to zero obtaining linear GG identities. So to obtain the same constriction at the order six we can derive three of the four graphs of the fourth order (one is linearly dependent by the relation itself)
and so on.

\section{Conclusion and Outlook}

In this paper we have used the method of the cavity fields via an interpolating parameter: we found this useful to derive an explicit expression for the free energy immediately below the critical temperature in terms of irreducible correlations functions of overlaps fluctuations (filled graphs). At the same time we obtained a way to generate systematically the constrictions to their free fluctuations (linear GG, \cite{5}\cite{11}) and we found that, in the broken replica phase, the SK behavior is restricted mainly by its internal symmetries more than by thermodynamic stability.
Future development should be finalized to apply this approach also to other mean field models with Gaussian quenched disorder and at the same time it should be useful to understand deeply which is the behavior of the tree of the constrictions at every non trivial order.

\bigskip

\bigskip

\textbf{Acknowledgments}

The author warmly thanks Francesco Guerra for a priceless scientific interchange. Truly thanks are also due to Peter Sollich, Pierluigi Contucci and Andrea Pagnani for useful conversations.

\bigskip

\


\addcontentsline{toc}{chapter}{Referencies}
\begin{thebibliography}{9}
\bibitem{1}D.Sherrington and S. Kirkpatrick

               A solvable model of spin glass.
               Phys. Rew. Lett. 35 (1975) 1792-1796
\bibitem{2}D.Sherrington and S. Kirkpatrick

               Infinite ranged models of spin glass.
               Phys. Rew. B17 (1978) 4384-4403
\bibitem{3}M. Mezard, G. Parisi, M.A. Virasoro

             Spin Glass Theory and Beyond
             World scientific publishing (1987)
\bibitem{4}L.A. Pasteur M.V. Shcherbina

                  The absence of self averaging of the order parameter in SK model.
                  J. Stat. Phys. 62 (1991) 1-19
\bibitem{5}S. Ghirlanda  F. Guerra

            General properties of overlap probability distributions in
            disordered spin systems.
            J. Phys. A, 31 (1998) 9149-9155
\bibitem{6}F. Guerra

            Sum rules for the free energy in the mean field spin glass model.
            Field institute comm. Vol. 30, 2001

\bibitem{7}E. Marinari G. Parisi J. Ruiz-Lorenzo  F. Ritort

             Numerical evidence for spontaneously broken replica symmetry in 3D spin glasses
             arXiv: cond-mat/ 950836v1
\bibitem{8}G. Parisi

             On the probabilistic formulation of the replica approach to spin glasses.
             arXiv: cond-mat/ 9801081v1
\bibitem{9}F. Guerra

                  The cavity method in the mean field spin glass model. Functional representation of the thermodynamic variables.
                  Advances in dynamical system and quantum physics. World Scientific
                  1995.
\bibitem{10}F. Guerra

              About the cavity fields in mean field spin glass models.
              arXiv: cond-mat/ 0307673v1
\bibitem{11}M. Aizenman P. Contucci

            On the stability of the Quenched State in Mean Field Spin Glass Models
            J. Stat. Phys. Vol.92 N.5-6 (1998) 765-783
\bibitem{12}G. Parisi

              Stochastic stability
              Disordered and complex system. AIP conf. proc.
\bibitem{13}M. Talagrand

                  Spin Glasses: A challenge for mathematicians.
                  Springer-Verlag, Berlin, 2003
\bibitem{14}F.Guerra F.L.Toninelli

              The thermodynamic limit in mean field spin glass models.
              Commun. Math. Phys. 230 (2002) 1,71-79
\bibitem{guerriero}F.Guerra

              Fluctuations and thermodynamic variables in Mean Field
              Glass.
              Stocastic Processes, Physics and Geometry, II.
              S.Albeverio at al. Eds. Singapore (1995).
\bibitem{baffioni}F. Baffioni, F. Rosati

              Some exact results on the ultrametric overlap distribution in mean field spin glass models.
              Eur. Phys. J. B 17, 439-447 (2000)


\end{thebibliography}
\end{document}